\documentclass[twocolumn,preprintnumbers,showpacs,amsmath,amssymb]{revtex4}

\usepackage{psfrag}
\usepackage{graphicx}
\usepackage{dcolumn}
\usepackage{bm}
\usepackage{epsfig}
\usepackage[notcite,notref]{}

\newcommand{ \sgn} {\operatorname{sgn}}
\newcommand { \sech} { \operatorname{sech}}

\begin{document}

\title{Exact Bogoliubov-de Gennes solutions for grey soliton backgrounds}

\author{P.B. Walczak}
\author{J.R. Anglin}
\email{janglin@physik.uni-kl.de}
\affiliation{
Physics Department and State Research Center OPTIMAS, University of Kaiserslautern,
Erwin-Schr\"odinger-Str. 46, D-67663 Kaiserslautern, Germany
}

\date{03.02.2011}

\begin{abstract}
We derive and discuss the complete set of exact solutions to the one-dimensional Bogoliubov-de Gennes equations for small amplitude excitations around general grey soliton solutions to the cubic nonlinear Schr\"odinger equation. Our results extend the previously known case of the motionless dark soliton background. We derive our non-zero frequency solutions using a variant of the factorization method for Schr\"odinger equations with reflectionless potentials. We also discuss the zero mode solutions at length.
\end{abstract}

\maketitle


\section{Introduction}

Solitons are a special class of solutions to a special class of nonlinear wave equations and appear as such in many different physical contexts. They are remarkable because they are, in a qualitative way, typical of a physical phenomenon so fundamental and ubiquitous that it is often entirely overlooked: The few parameters of a soliton, such as its position and velocity, stand out, as natural collective coordinates, from the infinitely many other degrees of freedom of the field in which the soliton propagates. Whenever physicists model something real as a finite idealized `system', we rely implicitly on the fact that a few particular degrees of freedom often do stand out in some way from among many. And whenever we speak of an open system, subject to noise or dissipation or even quantum decoherence, we are stating implicitly that the environment is somehow less prominent than the system. Prominence or its absence are thus very often taken for granted, which is more easily accepted than explained. It would be convenient to leave the issue to the philosophers, if it did not have such important physical consequences. It is worth looking at even simple models for prominence within physics. Solitons are a special case in which this general but elusive phenomenon can be clearly examined.

The first step towards understanding the dynamics of solitons, as preferred collective coordinates among infinitely many other field modes, is of course to identify the soliton solutions themselves. This has been done in very many cases, and the results constitute much of the extensive literature on solitons \cite{Zakharov1972,Zakharov1973,Kivshar1998,Carr2000}. The next step, however, is to learn about the solutions to the field equations within a configuration space neighborhood of the soliton, by solving the equations linearized about the soliton background. Far fewer explicit analytical results are known for this higher level question. The contribution of this paper is to provide them for a nontrivial one-parameter family of solitons studied in nonlinear optics, in hydrodynamics, and in Bose-condensed dilute gases -- the grey solitons of the cubic nonlinear Schr\"odinger equation (NLSE) in 1+1 dimensions (one dimension of space, one of time) \cite{Pitaevskii2003}.

Since grey solitons have been observed in quasi-one-dimensional Bose-Einstein condensates, where their motion has been confirmed \cite{Becker2008,Stellmer2008,Weller2008} to follow quite well the predictions of the NLSE in its role as the Gross-Pitaevskii mean field theory \cite{Busch2000}, there have been several recent theoretical studies which investigate solitons as more than isolated NLSE solutions. Various numerical methods have been applied to evolve many-body quantum states of a dilute Bose gas, and identified discrepancies between the quantum evolution and its mean field approximation given by the NLSE. This concerns the dark soliton dynamics in a lattice \cite{Mishmash2009,Mishmash2009a,Krutitsky2010}, or in a combined harmonic and lattice potential \cite{Martin2010,Martin2010a} also at finite temperatures. In \cite{Jackson2007,Jackson2007a,Cockburn2010} the dissipation of the soliton due to thermal fluctuations explained the experimentally observed increase of the soliton oscillation in the harmonic trap \cite{Burger1999}. Lately, \cite{Gangardt2010} used an effective Lagrangian for the dark soliton to estimate its decay and finite lifetime due to scattering on background fluctuations, which are treated in a Luttinger liquid approach.

The complementary work reported in this paper is in one respect much less ambitious than these quantum studies, and in another much more. We work entirely within a linearized theory in which there is not even any difference between quantum and classical dynamics, and so we ignore the entire issue of quantum corrections to the NLSE. On the other hand, though, our results are analytical, explicit, exact, and Hamiltonian. We therefore provide tools which may in future be used to understand the quantum dynamics that can be revealed, or at least suggested, by numerical many-body calculations.

The linearization of the NLSE about a stationary background is known in the dilute gas context as the Bogoliubov-de Gennes equation (BdGE). BdGE solutions are already known for the so-called `bright soliton' backgrounds \cite{Pethick2008} that are found for the case called `self-focusing' in optics, and `attractive' in gases \cite{Sinha2006}. Bright solitons are in a sense a trivial limit of solitons as collective degrees of freedom, however, because away from the soliton, the field amplitude falls off rapidly to zero. The bright soliton's position is thus simply a sort of center of mass for the entire field, and its motion is simply a Galilean boost of the entire system. A direct consequence is the gap in the Bogoliubov excitation spectrum, which simplifies the analysis of the soliton lifetime, for example, in contrast to the ungapped spectrum of the repulsively interacting gas \cite{Sacha2009}. 

\begin{figure}[h]
\psfrag{den}[b][r] {{ \small $| \psi_0|^2$}}
\psfrag{edg}[b][c] {$ \pm L$}
\psfrag{ori}[b][c] {$0$}
\psfrag{0}[c][t] {{ $0$}}
\psfrag{0.36}[c][c] {$ \beta^2$}
\psfrag{0.64}[c][c] {}
\psfrag{1}[c][c] {$ \mu$}
\includegraphics[width=8.6cm, angle=0]{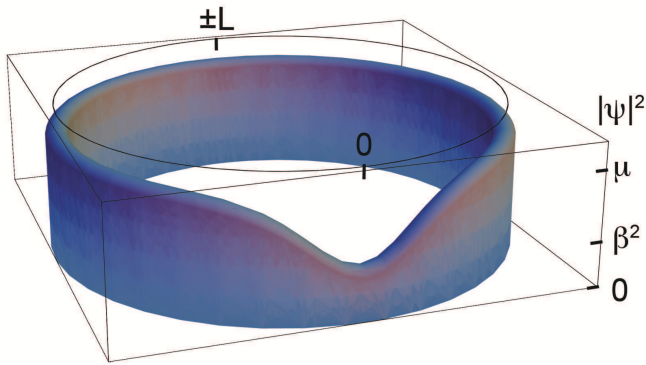}
\caption{A grey soliton located at $x=0$. Shown is $|\Psi|^2$ for periodic boundary conditions. The precise depth and breadth of the soliton `notch' are determined by the speed with which it moves. The Bogoliubov-de Gennes problem concerns the infinitely many modes of perturbations in $\Psi(x,t)$ around this background.}
\label{fig:density}
\end{figure}

With oppositely signed nonlinearity coefficent, the cubic NLSE is instead `self-defocusing' in optics, and represents repulsively interacting Bose-Einstein condensates. In this case it has `grey soliton' solutions, in which the field amplitude is locally reduced below its non-zero asymptotic value; there is a `notch' in the wave function modulus $|\Psi|$, as shown in Fig.~\ref{fig:density}, with a concomitant localized twist in the phase of $\Psi$. The grey soliton position is thus distinct from the background center of mass, and its velocity relative to the field at infinity is an independent degree of freedom from the global velocity of the entire system. BdGE solutions have recently been published for the strictly `dark' soliton, which does not move relative to infinity \cite{Dziarmaga2004,Negretti2008}. In this paper we present the BdGE solutions for grey solitons of all velocities in detail and at length, for periodic boundary conditions as well as on the infinite line. We exhibit a continuous spectrum of BdGE normal modes, plus discrete zero modes (normal modes of zero natural frequency) that are associated with perturbations of the soliton motion itself. We demonstrate the completeness of our modes as a basis for all BdGE solutions.

With the Bose-Einstein condensate foremost in mind as the physical realization of our equations, we analyze the BdGE as the canonical equations of motion of a Hamiltonian system, and discuss the amplitudes of our normal modes as canonical coordinates. This terminology should properly be modified for optical or hydrodynamic realizations of the NLSE and its linearization, since in these cases the variable we denote as $t$ is actually spatial distance, and so the functional we call the Hamiltonian generates translation instead of time evolution.  The periodic case that we refer to as a `ring geometry' would also rather apply, in optical contexts, to solutions in an unbounded spatial line but with a given temporal period. All our results remain formally valid, however, for any version of the BdGE equations. It is likewise unimportant for the present paper whether the dynamical system represented by the BdGE is quantum or classical. We are simply separating linear equations of motion by finding normal modes, and so our results are the same whether the amplitudes of those modes are operators or c-numbers.

The paper is organized as follows: Throughout the paper we will present results for both infinite and periodic systems, considering the two cases always in parallel. In order to provide a self-contained and pedagogical discussion of the problem, the first half of the paper is a review of standard material. In Section \ref{Hamiltonian} we review the NLSE as a Hamiltonian field theory, and in Section \ref{greysol} we describe the grey soliton solutions themselves. In Section \ref{bogoliubov} we review linearization around a background NLSE solution and present the BdGE as a linear Hamiltonian system. In Sections \ref{solutions} and \ref{sec zero} we finally present our main results, the explicit BdG eigenmodes for grey solitons of arbitrary velocity. A reader already familiar with both grey solitons and BdG theory might begin reading here. Section \ref{sec zero} is devoted to the zero modes, which are of special interest and subtlety. Even readers quite familiar with BdG problems may wish to consult Section \ref{bogoliubov} before Section \ref{sec zero}, in order to review the less commonly used canonical representation we adopt in this paper. Section \ref{discuss} discusses our results and suggests further lines of study. A series of appendices treats technical points skipped in the main text.


\section{Hamiltonian}
\label{Hamiltonian}
We begin with the Hamiltonian functional of a complex field $\Psi(x,t)$ in one spatial dimension: 
\begin{equation}\label{hamilton0}
H_0 = \int dx\,\left (\frac{1}{2}|\partial_x\Psi|^2 +\frac{1}{2}|\Psi|^4\right)\;.
\end{equation}
As dynamical variables in Hamiltonian field theory, $\Psi$ and $\Psi^*$ are symplectic co-ordinates \cite{Goldstein1980}; their real and imaginary parts correspond respectively to the usual canonical position and momentum fields. Consequently the canonical field equation for $\Psi$ derived from (\ref{hamilton0}) is the nonlinear Schr\"{o}dinger equation
\begin{equation}
\label{GP0}
i\partial_t\Psi = \frac{\delta H_0}{\delta\Psi^*}=-\frac{1}{2}\partial_{x}^2\Psi + |\Psi|^2\Psi.
\end{equation}
This equation may also be obtained as the Heisenberg equation of motion for the second-quantized destruction operator for a dilute Bose gas with repulsive particle interaction in the s-wave scattering limit, in appropriately re-scaled dimensionless variables, with the quantum field $\hat{\Psi}$ then replaced with the c-number field $\Psi$ in Gross-Pitaevskii mean field theory. Nevertheless, it is worth appreciating that Eqn.~\eqref{GP0} is also a perfectly good classical Hamiltonian field equation, and appears as such in several physical contexts.

Our goal is to describe the linearized dynamics generated by $H_0$ around a time-independent solution to the field equation -- namely, a soliton solution. But in fact we can take advantage of two symmetries of $H_0$, and obtain linearized dynamics around solutions with generalized time-independence, as follows. If we define a new field $\psi(x,t)$ such that
\begin{equation}
\label{convertPsi}
\Psi(x,t) = e^{-i (\mu+v \beta-v^2/2) t}\psi(x-(\beta-v) t,t)
\end{equation}
for constants $\beta$, $v$ and $\mu$, we can describe any field $\Psi$ in terms of $\psi$ instead.

The field equation for $\psi$ is derived straightforwardly from \eqref{GP0}, and after changing to a moving co-ordinate 
frame $ (x', t) \equiv (x- (\beta-v) t,t)$ reads
\begin{equation}
\label{GP1}
i\partial_{t} \psi(x', t) = -\frac{1}{2} \partial_{x'}^2\psi + i (\beta-v) \partial_{x'} \psi + |\psi|^2\psi - \tilde{\mu}\psi \;,
\end{equation}
where we have defined
\begin{equation}
\label{chemical}
\tilde{\mu} \equiv \mu+v \beta-\frac{v^2}{2}
\end{equation}
to shorten the equations. Throughout this paper we will work in this moving frame and drop the $'$ to simplify notation. The corresponding Hamiltonian functional of $\psi$, whose symplectic field equation is \eqref{GP1} just as that of $H_0$ is \eqref{GP0}, is
\begin{equation}
\label{hamilton}
\begin{split}
H &= \int dx\,\left(\frac{1}{2}|\partial_x\psi|^2 +\frac{1}{2}(|\psi|^2- \tilde{\mu})^2\right. \\
  &\qquad \qquad+\left.i\frac{\beta-v}{2}[\psi^*\partial_x\psi - (\partial_x\psi^*)\psi]   \right).
\end{split}
\end{equation}
The differences between $H$ and $H_0$ are the constants of motion associated with the symmetries of phase and position translation, and the change from $\Psi$ and $H_0$ to $\psi$ and $H$ is a canonical transformation. In terms of a Bose gas, the operators at issue here are the total particle number and total linear momentum, which commute with the Hamiltonian. Evolution within the subspace of definite number and momentum may be generated equivalently by the quantized versions of $H$ and $H_0$, since the discrepancy between them is only an unobservable phase prefactor mulitplying all amplitudes in the subspace.

The main result of this paper will be to explicitly diagonalize the linearization of $H$ around a non-trivial class of time-independent solutions to \eqref{GP1}: grey solitons.


\section{Grey solitons}
\label{greysol}
The grey soliton \cite{Tsuzuki1971,Busch2000} with position $x_0$ (constant in our co-moving frame), and constant velocity $\beta$ relative to the background phase gradient, is
\begin{equation}
\label{sol}
\begin{split}
\psi_0(x) &= e^{i \theta} e^{-i v (x-x_0)} \{ i \beta + \kappa \tanh \kappa (x-x_0) \}\\
\kappa &=\sqrt{\mu -\beta^2}\;.
\end{split}
\end{equation}
The constant overall phase $ \theta$ can be chosen to be zero without loss of generality. For $\kappa |x-x_0|\gg 1$, $\psi$ has the constant phase gradient $v$ corresponding in the BEC context to a constant gas velocity $ v$. We will sometimes refer to this solution parameter as the (constant) gas velocity in the following. The soliton itself is moving relative to our original frame slower by $\beta$ (or faster by $-\beta$) than the background gas motion. It is thus important to recognize that $\beta$ and $v$ are independent parameters. A grey soliton moves with the phase gradient of its background field, but it is also able to move independently of its surrounding field, by deforming its shape. Note that the relative soliton speed $\beta$ is restricted to lie within the range 
\begin{equation}
\label{condbeta}
|\beta| \leq \sqrt{\mu} = c \, ,
\end{equation}
i.e.~the soliton speed can never exceed the speed of sound $c$ of the background gas. For the extremal values of $\beta$, the grey soliton reduces to a translationally invariant $\psi$ with constant modulus and phase gradient.

The squared modulus $ |\psi(x)|^2$ (corresponding to condensate linear density, or light intensity in a nonlinear fiber) has a minimum at $x=x_0$, where it drops to 
\begin{equation}
|\psi_0|^2_{min} = \beta^2.
\end{equation}
The case $\beta=0$ is therefore called the `dark soliton' (since in fiber optics it represents a perfectly dark spot) although grey solitons with nonzero $\beta$ are also sometimes referred to as dark. $|\psi_0|^2$ rises with $|x-x_0|$ toward its asymptotic value of $ \kappa^2 + \beta^2$. Except for the extremal values of $\beta$, for which $\kappa$ becomes very small, grey solitons are well localized objects. The asymptotic value of $|\psi_0|^2$ is nearly reached already by around $\kappa |x-x_0|\gtrsim2$, and the remaining deficit $\kappa^2+\beta^2-|\psi_0|^2$ then decays exponentially on the soliton length scale $1/\kappa$, see Fig.~\ref{fig:density}.

The soliton solution can also be characterized by its phase behavior. Across the interval $|x-x_0|<d$, $\psi$ has a total phase difference 
\begin{equation}
\label{phase}
\Delta \phi_d = \pm \pi- 2 \arctan \left ( \frac{\beta}{ \kappa \tanh \kappa d}\right) +2 v d
\end{equation}
for $\beta \gtrless 0$. Apart from phase change due to the gas velocity $v$, the largest jump in phase occurs over $\kappa d \sim 2$. The grey soliton is localized in its phase behavior as well as in its modulus.

A grey soliton with larger $\beta$ is moving faster relative to its background, but it is a longer and shallower disturbance of the background. For this reason it has lower energy $H$ than a dark soliton with $\beta=0$, which does not move relative to its background at all. Consequently one may state that dark solitons are energetically unstable to acceleration \cite{Busch2000}!

Periodic boundary conditions may also be imposed on our system; in the BEC context, for example, the gas may be confined in a tight toroidal trap \cite{Kuga1997,Morizot2006,Ryu2007a}. The soliton solution \eqref{sol} is in this case no longer exact, because it cannot be made smoothly periodic; the exact periodic solutions may be expressed in terms of Jacobi elliptic functions \cite{Carr2000}, and approximate solutions can be obtained \cite{Smyrnakis2010}. In this paper we will nonetheless retain certain cases of \eqref{sol} as our solutions, even in the ring geometry, by assuming that the ring circumference $2L$ is much larger than the soliton size $1/\kappa$. In this limit the unavoidable discontinuity in $\partial_{x}\psi$ at $x-x_{0}=\pm L$ is only of order $ \mathcal{O}(e^{- 2 \kappa L})\lll 1$, which we consider negligible.  (It should therefore be noted that for fixed system size $2L$, we do not treat cases with $\kappa$ arbitrarily close to zero; but we do treat any fixed $\kappa$, for sufficiently large $L$.) Throughout this paper, whenever we discuss the case with periodic boundary conditions, we will present equations that are only strictly valid up to corrections of this order, without further comment. 

The grey soliton solution is continuous, and furthermore periodic on the ring, if the phase difference across the whole system is $ \Delta \phi_L = 2m \pi$ with integer $m$. Choosing the chemical potential $ \mu$ and the relative soliton speed $ \beta$ as the two independent parameters determines the allowed values of the gas velocity 
\begin{equation}
\label{velocity}
v = \frac{2 \pi m - \pi}{2L} + \frac{1}{L} \arctan \frac{\beta}{ \kappa} \, .
\end{equation}
It follows that $v$ has discrete values, and for finite $L$, the gas velocity cannot actually be zero for any $|\beta|< c$, but has a minimum value of order $1/L$. Since the resulting phase gradient of $\psi_0$ is constant over most of the entire $2L$ circumference, it makes a finite contribution to the total angular momentum. For some purposes this must be borne in mind, though for others a $v$ of order $1/L$ will be entirely negligible. We may still consider $v$ arbitrarily close to any given value, for sufficiently large $L$.


\section{Linearization of the GPE}
\label{bogoliubov}
We can obtain a larger class of solutions to the GPE than \eqref{sol} alone, in the form of solutions that are close to it: 
\begin{equation}
\label{near} 
	\psi(x,t) = \psi_{0}(x)+ \delta\psi(x,t)
\end{equation}
with $\delta\psi$ being small. In the Bose gas context, an operator-valued version of this perturbation $\delta\hat{\psi}$ is used to describe the leading quantum corrections to the Gross-Pitaevskii mean field theory. Within the linearized dynamics we will discuss, however, the quantization of $\delta\psi$ is trivial, being that of a non-relativistic free field. We will therefore continue classically, analyzing the linearized Hamiltonian field theory of $\delta\psi(x,t)$. Application to the quantum problem can be achieved with literally no greater change than simply placing operator accents $\hat{\ }$ over all of our classical canonical variables.

Inserting \eqref{near} into our Hamiltonian \eqref{hamilton}, we find that terms linear in $\delta\psi$ vanish because $\psi_{0}$ satisfies the time-independent GPE. The second order term $H_{2}$ in $H$ is therefore leading, and we ignore all higher order terms in this paper:
\begin{equation}
\label{second}
\begin{split}
H_2 &= \frac{1}{2} \int \! dx \, \delta \psi^{ \ast} \bigg[\! \left(\! - \frac{1}{2} \partial_x^2+ 2 | \psi_0|^2- \tilde{\mu} \right) \!\delta \psi \\
& \hspace{2.2cm}+ \psi_0^2 \delta \psi^{ \ast} + i (\beta-v) \partial_x \delta \psi \bigg] \!+ \mathrm{c.c.}
\end{split}
\end{equation}
The canonical equation of motion derived from $H_{2}$ for $\delta\psi$ co-incides exactly with the linearization of the GPE about $\psi_{0}$:
\begin{equation}
\label{BdG}
i \partial_t \delta \psi=\! \left (\! -\frac{1}{2} \partial_x^2 +i (\beta-v) \partial_x + 2 | \psi_0|^2 -\tilde{\mu} \! \right)\! \delta \psi + \psi_0^2 \delta \psi^{ \ast}.
\end{equation}
This is the time-dependent Bogoliubov-de Gennes equation (BdGE). Solving it in terms of temporally harmonic normal modes is equivalent to diagonalizing $H_{2}$ by a canonical transformation. Our contribution in this paper is to do so explicitly and exactly for soliton background $\psi_{0}$ for all values of $\beta$. This expands upon the previously known solution for $\beta = 0$ \cite{Dziarmaga2004,Negretti2008}, and significantly extends the set of backgrounds for which explicit BdGE solutions are known. In this section we set up the problem in detail, and discuss the orthonormality and completeness of BdGE normal modes, before presenting our explicit solutions in the next sections.

\subsection{Canonical normal modes}
We begin by expanding
\begin{equation}
\label{dev}
 \delta \psi (x,t) = \sum_{ \eta} (R_ \eta(x,t) q_\eta(t) \! + \! i S_ \eta(x,t) p_\eta(t))
\end{equation}
for a set of real-valued classical canonical variables $ \{ q_{ \eta}, p_{ \eta} \}$, labelled with index $\eta$. The summation over $ \eta$ in \eqref{dev} is used in the symbolic manner. For the finite system it is a true summation over discrete values of $ \eta$, whereas for an infinite system it stands for the summation over discrete values and an integration over continuous values of $ \eta$. 

Eqn.~\eqref{dev} defines $R_{\eta},S_{\eta},q_{\eta},p_{\eta}$, given $\delta\psi$. It is to be understood as a mapping from one set of infinitely many degrees of freedom, namely the real and imaginary parts of $\psi$ for each position $x$, to another set, namely the pairs $q_{\eta},p_{\eta}$ for each $\eta$. In order to be sure that we have fully solved the evolution of $\delta\psi$, we must require that this mapping be complete; and we would additionally like the mapping to be one-to-one. This means that every possible $\psi$ can be obtained with one and only one set of $\{q_{\eta},p_{\eta}\}$. Hence, \eqref{dev} should be invertible, and for any instant $t$ we can regard each $q_{\eta}$ and $p_{\eta}$ as a functional of $\psi(x)$.  

Since canonical co-ordinates are important for many applications (including quantization), and since any solution of a linear dynamical system can be expressed very easily in terms of phase space co-ordinates, we can finally demand that the set $\{q_{\eta},p_{\eta}\}$ be canonical variables. Consequently, they fulfill the fundamental Poisson bracket relations
 \begin{align}
 \label{funpoisson}
 [q_{ \eta}, p_{ \eta^{ \prime}}] &= \delta_{\eta \eta^{ \prime}} &  [p_{ \eta}, p_{ \eta^{ \prime}}] &= [q_{ \eta}, q_{ \eta^{ \prime}}] =0 \; ,
 \end{align}
where the Poisson bracket is defined as
\begin{equation}
\label{poissonpsi}
	[A,B] \equiv-i \! \int\! dx \left(\frac{\delta A}{\delta\psi(x)}\frac{\delta B}{\delta\psi^{*}(x)}-\frac{\delta B}{\delta\psi(x)}\frac{\delta A}{\delta\psi^ \ast(x)} \right)
\end{equation}
for any functionals $A,B$. (Throughout this paper the delta with mode indices is a Kronecker delta for discrete degrees of freedom and has to be understood as a Dirac delta distribution for variables that depend on a continuous index.) 

The definition \eqref{poissonpsi} immediately implies that
\begin{equation}
 \label{complete}
 \begin{split}
[ \delta \psi (x), \delta \psi ^{ \ast} ( x ^{ \prime})] &=-i \delta (x- x ^{ \prime}) \\
[ \delta \psi (x), \delta \psi ( x ^{ \prime}) ] &= 0\,.
 \end{split}
 \end{equation}
Inserting \eqref{dev} into \eqref{complete} and applying \eqref{funpoisson} yields the completeness relations (or partition of unity)
 \begin{equation}
\label{completerel}
\begin{split}
\sum_{ \eta} ( R_ \eta (x) S_ \eta ^{ \ast} (y) + S_ \eta (x) R_ \eta ^{ \ast} (y)) & = \delta (x- y)\\
\sum_{ \eta} ( R_ \eta (x) S_ \eta (y) - S_ \eta (x) R_ \eta (y)) & = 0 \; .
\end{split}
\end{equation}
These latter must be checked to confirm that all normal modes have been found, and none was overlooked. 

We show in Appendix \ref{proof}, without any reference to the BdG equations, that invertibility and canonicity of \eqref{dev} imply the orthonormality relations
 \begin{equation}
\label{ortho}
\begin{split}
\int dx \left( R_ \eta S_{ \eta^{ \prime}}^{ \ast} + R_ \eta^{ \ast} S_{ \eta^{ \prime}} \right) &= \delta_{\eta \eta^{ \prime}}\\
 \int dx \left( S_ \eta S_{ \eta^{ \prime}}^{ \ast} - S_ \eta^{ \ast} S_{ \eta^{ \prime}} \right) &= 0\\
\int dx \left( R_ \eta R_{ \eta^{ \prime}}^{ \ast} - R_ \eta^{ \ast} R_{ \eta^{ \prime}} \right) &= 0\;,
\end{split}
\end{equation}
which in turn imply that the Poisson bracket is equivalently given as
\begin{equation}
\label{defpoisson}
[ A,B]= \sum_ \eta  \left( \frac{ \partial A}{ \partial q_{ \eta}} \frac{ \partial B}{ \partial p_{ \eta}}-\frac{ \partial A}{ \partial p_{ \eta}}\frac{ \partial B}{ \partial q_{ \eta}} \right)\;,
\end{equation}
as indeed it may be given in terms of any set of canonical variables. This means that orthonormality of BdG modes does not need to be understood as a fortunate co-incidence due to the particular form of the BdGE, but can be considered a general consequence of Hamiltonian dynamics, which the BdGE of course obey. And neither the completeness nor the orthonormality relations are consequences of quantum mechanics; the correspondence of classical Poisson brackets, and quantum commutators, is exact, with a factor of $i\hbar$. Both these relations stem simply from Hamiltonian mechanics, which is common to both classical and quantum physics.
It follows from the BdGE \eqref{bdgmode}, though, that the inner product in \eqref{ortho} automatically vanishes, for all cases where $\Omega_{\eta}^{2} \neq \Omega_{\eta'}^{2}$. This will be addressed in Appendix \ref{orthonormality}.

\subsection{Time-independent Bogoliubov-de Gennes Equations}

We now turn to the particular form of the BdGE equations. Anticipating that the normal modes of a linear system evolve as harmonic oscillators, we assume that the time evolution of the canonical variables is derived from a Hamiltonian of the form
\begin{equation}
\label{hamfunction}
H_2= \frac{1}{2} \sum_{ \eta} \left( \frac{p_{ \eta}^2}{m_ \eta} + m_ \eta \Omega_{ \eta}^2 q_{ \eta}^2 \right)
\end{equation}
in which $m_{\eta}$ is a mass which will be dimensionless in our chosen units, and $\Omega_{\eta}$ is the eigenfrequency of the normal mode labelled by $\eta$. The Hamiltonian evolution generated by this form of $H_{2}$ implies a set of linear equations among the $R_{\eta}(x),S_{\eta}(x)$, namely the time-independent BdGE:
\begin{equation}
\label{bdgmode}
\begin{split}
 m_\eta \Omega_{ \eta}^2 S_ \eta(x) &=  H_B R_ \eta (x)+ \psi_0^2 R_ \eta ^{ \ast}(x) \\
 m_\eta^{-1} R_ \eta(x)  &=  H_B  S_ \eta (x)- \psi_0^2 S_ \eta ^{ \ast}(x)
\end{split}
\end{equation}
with the differential operator
 \begin{equation}
 \label{hb}
H_B  \equiv -\frac{1}{2} \partial_x^2 +i (\beta-v) \partial_x + 2 | \psi_0|^2 - \tilde{\mu} \; .
\end{equation}
Inserting these eigensolutions $R_{\eta},S_{\eta}$ into \eqref{dev} and \eqref{second} will reveal that $H_{2}$ as given by \eqref{second} is indeed equal to \eqref{hamfunction}. This can be seen simply by applying \eqref{bdgmode} and the orthonormality properties, and integrating by parts, without needing to use the explicit forms of $R_{\eta},S_{\eta}$ that we supply below. This confirms that the $R_{\eta},S_{\eta}$ which solve \eqref{bdgmode} are those which diagonalize $H_{2}$, and that the solutions to the time-dependent BdGE \eqref{BdG} are indeed of the form \eqref{dev}, with the explicit time dependence
\begin{equation}
\label{BdGsol}
\begin{split}
	q_{\eta}(t) &= q_{\eta}(0)\cos\Omega_{\eta}t + \frac{p_{\eta}(0)}{m_{\eta}\Omega_{\eta}}\sin\Omega_{\eta}t \\
p_{\eta}(t) &= p_{\eta}(0)\cos\Omega_{\eta}t - m_{\eta}\Omega_{\eta}q_{\eta}(0)\sin\Omega_{\eta}t\;.
\end{split}
\end{equation}
Cases with $\Omega_{\eta}\to0$ are covered here simply by taking the limit.

\subsection{Normal mode spectrum}

It is possible in general for Bogoliubov eigenfrequencies $\Omega_{\eta}$ to be complex. In this case the system is dynamically instable, and although it implies that the linearization approximation must break down on logarithmically short time scales, within that time frame it is by no means unphysical. The case of complex $\Omega_{\eta}$ requires several significant generalizations to the Bogoliubov-de Gennes theory for real $\Omega_{\eta}$, but in the problems treated in this paper, it turns out that all $\Omega_{\eta}$ are indeed real. We prove this simply by demonstrating that the set of modes with real $\Omega_{\eta}$ is complete in the sense of \eqref{completerel}. We can therefore discuss the nature of the BdG normal mode spectrum without treating the extra complications that arise for complex $\Omega_{\eta}$.

Since \eqref{bdgmode} is a coupled system of two complex second order differential equations, there are in principle eight linearly independent solutions for each eigenvalue $ \Omega_ \eta$.  The overall normalization of any solution is fixed, up to a phase, by the orthonormality condition \eqref{ortho}. The phase is arbitrary, but in a slightly non-trivial way. Because of the explicit complex conjugates appearing in the equation, rescaling a solution $(R_{\eta},S_{\eta})\to e^{i\alpha} (R_{\eta},S_{\eta})$ does not in general provide another solution. The phase freedom that does exist is rather 
\begin{equation}
\begin{split}
R_{\eta}&\to R_{\eta}\cos\alpha-i m_\eta \Omega_{\eta}S_{\eta}\sin\alpha \\
S_{\eta}&\to S_{\eta}\cos\alpha-i\frac{R_{\eta}}{ m_\eta \Omega_{\eta}}\sin\alpha\;.
\end{split}
\end{equation}
Considering \eqref{dev} and \eqref{hamfunction}, it is easy to see that the second solution merely reproduces exactly the same $\delta\psi$ as the first, but with the co-efficient relabeling 
\begin{equation}
\begin{split}
	q_{\eta} & \to q_{\eta}\cos\alpha +\frac{p_{\eta}}{m_\eta \Omega_{\eta}}\sin\alpha\\
	p_{\eta}&\to p_{\eta}\cos\alpha - m_\eta \Omega_{\eta}q_{\eta}\sin\alpha\;.
\end{split}
\end{equation}
This relabeling happens to be a canonical transformation; but this does not change the fact that these two solutions do not represent distinct degrees of freedom in $\delta\psi$, but merely a double counting of the same configuration. Eliminating such double counting by picking one value of $\alpha$ arbitrarily is equivalent to discarding half of the eight solutions, namely those given by $\alpha+\pi/2$. There are thus only four distinct canonical solutions for each $\Omega_{\eta}$. Boundary conditions such as square-integrability or periodicity will generally eliminate at least two of these, providing zero, one, or two normal modes per non-zero $\Omega_{\eta}$.

A further symmetry is relating solutions with different $m_{\eta}$. By canonical rescaling of $q_{\eta},p_{\eta}$ it is possible to set $|m_{\eta}|\to 1$ without loss of generality -- and for the rest of this paper we will assume that this has already been done. Please note that it is not possible to change the sign of $m_{\eta}$. Both signs are possible, and both must be considered for any $\Omega_{\eta}$, because if $(R_{\eta},S_{\eta})$ is a solution to \eqref{bdgmode}, then $(R_{\eta},-S_{\eta})$ is a solution for $m_{\eta}\to - m_{\eta}$.  Examining \eqref{dev}, however, we see that these two solutions correspond to exactly the same configuration of $\delta\psi$, with the trivial co-ordinate relabeling $p_{\eta}\to -p_{\eta}$. They therefore represent the same degree of freedom, counted twice, and we not only can, but must, discard one and retain only the other. Inspection of  \eqref{ortho} shows that the co-efficients $q_{\eta},p_{\eta}$ will be canonical for one and only one solution in each of these pairs, since $[q,-p]=-[q,p]$ trivially. So we can maintain canonicity by always choosing the canonical co-efficients to retain. 

This means that the sign of $m_{\eta}$ for each mode is just as much a physical fact as its eigenfrequency $\Omega_{\eta}$. The sign of $\Omega_{\eta}$ itself has no meaning whatever; it has no effect in either the time-independent BdGE \eqref{bdgmode} or the time-dependent solution \eqref{BdGsol}. $\Omega_{\eta}$ may therefore be taken as positive without loss of generality. Inspecting $H_{2}$ in \eqref{hamfunction}, one sees that it is the sign of $m_{\eta}$ which determines whether the energy contribution from exciting a given mode is positive or negative. A negative mass simply means an energetic instability; an excitation of this mode lowers the energy. The existence of such modes is of course incompatible with strict thermodynamic equilibrium, but it is perfectly physical for perturbations around an excited background such as our grey soliton, which in many systems may nonetheless be very long-lived. 

We remark finally that in many BdG cases the index $ \eta$ can immediately be identified with a Fourier wave number $k$, but in a problem which is not translational invariant such as ours, this is in general not so. However, in our soliton problem there is an asymptotic translational invariance, and most of our modes will indeed have a Fourier index $\eta\to k$, such that for $\kappa |x-x_{0}|\gg 1$ the behave as sinusoidal functions of $kx$. But as will be shown later, there will be one extra mode of frequency zero. For this mode we will use the single additional index $ \eta =z$. 

\subsection{Comparison with complex co-ordinates}
In the literature on the dilute Bose gas in particular it is more common to employ a different form of expansion from \eqref{dev}, which is, however, fully equivalent to it. This simply uses complex variables $ a_\eta, a_ \eta ^ \ast $ instead of the real pairs $q_{\eta},p_{\eta}$:
\begin{equation}
\label{coherent}
\delta \psi = \sum_{ \eta} ( u_{ \eta} (x,t) a_{ \eta}(t) + v_{ \eta} ^{ \ast} (x,t) a_{ \eta} ^{ \ast}(t))\; .
\end{equation}
For each individual non-zero frequency the transformation between the sets is defined by
\begin{equation}
\label{symplectic}
\begin{bmatrix} a_{ \eta} \\ a_{ \eta} ^{ \ast} \end{bmatrix} = \frac{1}{ \sqrt{2}} \left( \sqrt{ \Omega_{ \eta}} \begin{bmatrix} q_{ \eta} \\ q_{ \eta} \end{bmatrix} + \frac{i}{ \sqrt{ \Omega_{ \eta}}} \begin{bmatrix} p_{ \eta} \\ - p_{ \eta} \end{bmatrix} \right ) \; .
\end{equation}
The two variable sets are connected by a so-called extended canonical transformation \cite{Goldstein1980} with a scaling parameter $ \lambda = - i$. In the quantum version of this problem, just as the $\psi$ field becomes a destruction operator, and our phase space co-ordinates $q_{\eta},p_{\eta}$ become operators $\hat{q}_{\eta},\hat{p}_{\eta}$ with canonical commutation relations, these complex co-ordinates $a_{\eta},a^{*}_{\eta}$ become annihilation and creation operators $\hat{a}_{\eta},\hat{a}^{\dagger}_{\eta}$ for quasi-particles. In this sense there is nothing particularly quantum mechanical about the Bogoliubov expansion \eqref{coherent}, any more than there is about \eqref{dev}. They are simply equivalent alternative representations of linear Hamiltonian dynamics.

Expressing the non-zero frequency part of \eqref{hamfunction} in terms of the complex variables yields the familiar Hamiltonian
\begin{equation}
\label{hama}
H_2= \sum_{\eta} \Omega_{ \eta} a_{ \eta} ^{ \ast} (t) a_{ \eta} (t) \; .
\end{equation}
In analogy to \eqref{bdgmode}, the resulting linear equations in the complex co-ordinate representation are
\begin{equation}
\label{eigenval}
\begin{split}
\Omega_\eta u_\eta  &=  H_B  u_\eta + \psi_0^2 v_\eta \\
\Omega_\eta v_\eta  &= - \psi_0^{ \ast 2} u_\eta -H_B ^ \ast  v_\eta \, ,
\end{split}
\end{equation}
with the same differential operator $H_B$ defined in \eqref{hb}. This is a form of the BdGE nowadays widely used in the literature. In the $a,a^{*}$ formalism it is conventional to make the sign of $\Omega_{\eta}$ meaningful and eliminate $m_{\eta}$ by absorbing $m_{\eta} = \pm 1$ in $\Omega_{\eta}$.

In this paper we will mainly use the representation with the set $ \{ q_\eta, p_\eta \}$, because although it is in some respects more cumbersome, it has one important advantage: since the Hamiltonian \eqref{hamfunction} does not vanish for zero frequency, unless $m_{\eta}$ happens also to be infinite, there is no ambiguity or singularity in the $\{q_{\eta},p_{\eta}\}$ expansion when the mode frequency $\Omega_{\eta}\to 0$. In the $\{a_{\eta},a_{\eta}^{*}\}$ representation, in contrast, a mode for which $\Omega_{\eta}$ vanishes must either have its Hamiltonian exactly zero, or else have some singularity in \eqref{coherent}. To avoid such complications we retain the $\{q_{\eta},p_{\eta}\}$ representation. We will be devoting considerable discussion to the modes with zero frequency, both because their technical subtleties require it, and because one of them is of particular interest, in that it represents the motion of the soliton itself.

We are now ready to present the main result of this paper: the explicit solution of \eqref{bdgmode} for a complete set of $\Omega_{\eta}, m_{\eta}$.

\section{Solutions of the Bogoliubov-de Gennes equation}
\label{solutions}

\subsection{Non-zero frequency solutions}
\label{phonon}
In this subsection we present the explicit non-zero frequency solutions of the BdGE (for which $\eta\to k$) for the general case of a grey soliton; their derivation through a factorization technique analogous to Schr\"odinger supersymmetry is shown in the next subsection. The special case of the zero frequency solutions follows further below. 

If we rescale all the positive masses to $m_k =1$, the solutions to the Bogoliubov equations \eqref{bdgmode} for non-zero frequency in the co-moving frame in case of both finite and infinite system can be expressed as
\begin{align}
\label{littler} R_k(\tilde{x}) &= e^{-i v \tilde{x}} \sqrt{ \Omega_{ \eta}} \left( \mathrm{Re} \, r_k ( \tilde{x}) + i \mathrm{Im} \, s_k ( \tilde{x})   \right) \\ 
\label{littles} S_k (\tilde{x})&=\frac{e^{-i v  \tilde{x}}}{ \sqrt{ \Omega_{ \eta}}} \left( \mathrm{Re} \, s_k ( \tilde{x})   + i \mathrm{Im} \, r_k ( \tilde{x}) \right), 
\end{align}
where we have introduced the abbreviation $ \tilde{x} \equiv x- x_0$. The functions $r_k, s_k$ are given by
\begin{align}
\label{contsolr} r_k(x)  &= \frac{ \mathrm{N}_k}{  \Omega_k} e^{i k x}\bigg( \frac{k ^3}{2} - 2 \beta \Omega_k + i k^2 \kappa \tanh \kappa x\nonumber\\
&\qquad\qquad\qquad\qquad + k \kappa^2 \! \sech^2 \! \kappa x \bigg) \\
\label{contsols} s_k(x) &= \mathrm{N}_k\,  e^{i k x}(k + i 2 \kappa \tanh \kappa x) \; 
\end{align}
with the mode frequencies $\Omega_k$ being
\begin{equation}
\label{bogfre} \Omega_k + \beta k = \sqrt{\frac{k^4}{4} + c^2 k^2} \; .
\end{equation}
This represents exactly the same dispersion relation as for a constant background solution $ \psi_{ \pm}= e^{-i v  \tilde{x}} ( i \beta \pm \kappa)$ equal to the grey soliton background solution in the limit $x \to \pm \infty$. For a dark soliton ($\beta=0$) this has already been demonstrated in \cite{Dziarmaga2004,Negretti2008}. 

The normalization constant $ \mathrm{N}_k$ that ensures \eqref{ortho} is
\begin{equation}
\label{defnorm}
\frac{1}{\mathrm{N}_k^2} \stackrel{ \mathrm{\delta}}{=} 8 \pi \left( \Omega_k + \beta k\right) \end{equation}
for delta-function normalization in the infinite domain, with a continuous spectrum of $k$, and
\begin{equation}\label{defnorm2}
\frac{1}{\mathrm{N}_k^2} \stackrel{ \mathrm{ring}}{=} 8 L ( \Omega_k + \beta k)-4 \frac{ \kappa k^2}{ \Omega_k}
\end{equation}
for the finite ring with its discrete spectrum. For the demonstration of the orthonormality of these modes, see Appendix \ref{orthonormality}.

For the finite ring geometry it is also necessary to select the discrete set of $k$ for which the solutions are periodic (up to exponentially small corrections, as discussed above). Just as with the background solution $\psi_{0}$, these are obtained by imposing continuity for the $\delta\psi$ modes and their first derivatives. As long as $\psi_{0}$ is thus periodic, the condition for periodicity of $\delta\psi$ reduces to
\begin{equation}
\label{conditionk}
k\cot k L=2\left( \kappa-\frac{\beta \Omega_k}{\kappa k} \right ) \; .
\end{equation}
In case of the dark soliton $ \beta =0$, this condition is equivalent to the one given in \cite{Negretti2008}.

From the explicit solutions above one can see that any perturbation of the grey soliton background made from the normal mode spectrum passes through the soliton without being reflected, which is expected for the integrable model \cite{Muryshev2002,Gangardt2010}.
 
The proof that the above non-zero-frequency modes together with the zero frequency modes described in Section \ref{sec zero} below are a complete set in the sense of \eqref{completerel} is given in Appendix \ref{completeness}. It proceeds simply by direct integration, which can be performed analytically, but is somewhat involved. It is not hard to see that no additional finite-frequency modes are expected, however. Due to the symmetry of the Hamiltonian for non-zero frequency modes mentioned in \ref{bogoliubov}, it is only necessary to find four independent solutions for a given eigenfrequency. For non-zero frequency $\Omega$, the Bogoliubov dispersion relation \eqref{bogfre} is a quartic equation in the mode index $k$. Assuming all frequencies are real and that  $\kappa^2 \neq 0$, this equation must in general have two real roots and two mutually complex conjugate roots for $k$ \cite{Teubner1996}. Since the complex $k$ solutions are un-normalizable in the infinite domain, and can be shown incompatible with periodicity on the ring, we are left with two distinct solutions with real $k$ for each $\Omega$, and these are the solutions we have displayed.

\subsection{Comparison with constant background solutions}
\label{constant}
It is quite instructive to redo the complete linearization procedure for the constant background solution $ \psi_{\pm}= e^{-i v  \tilde{x}} ( i \beta \pm \kappa)$, which is equal to the soliton background solution $ \psi_0$ at $\tilde{x} \to \infty$ or $\tilde{x} \to - \infty$. As mentioned before, one obtains the same dispersion relation \eqref{bogfre}, but instead of \eqref{contsolr} and \eqref{contsols} the solutions are given by
\begin{align}
r_k^{ \pm} (x)&= e^{i k \tilde{x}} \frac{ \mathrm{N}_k}{ \Omega_k} \left( \frac{k^3}{2}- 2 \beta \Omega_k \pm i k^2 \kappa \right)\\
s_k^{ \pm} (x)&= e^{i k \tilde{x}} \mathrm{N}_k (k \pm i 2 \kappa) \; .
\end{align}
The normalization constant of these non-zero frequency solutions is the same for the infinite and finite system and is given by \eqref{defnorm}. Except for the phase factors the solutions for the BdGE for constant background are constants. Comparing these with \eqref{contsolr} and \eqref{contsols} one immediately sees that solutions for the constant background $ \psi_{ \pm}$ coincide with the soliton solutions at $ \tilde{x} \to \pm \infty$.

\subsection{Derivation of non-zero frequency solutions via factorization}
\label{suzie}
In one dimensional linear Schr\"odinger systems, scattering off a potential can sometimes be simplified with a factorization of the Hamiltonian. Factorization is a technique which relates a problem at hand to a system with a simpler potential by using a proper superpotential. In this way one can easily solve the one dimensional scattering due to a potential of the form $ \kappa^2 \sech^2 \! \kappa x$, for example, since it is connected to a constant potential \cite{Boya1988}. We have demonstrated in the preceding subsection that the grey soliton background acts like a reflectionless potential for the non-zero frequency modes just like the $ \kappa^2 \sech^2 \! \kappa x$-potential, which suggests the usage of a method similar to the factorization method. To do so, we are going to rewrite the BdGE \eqref{bdgmode} in terms of the functions $r_k, s_k$, related to the $R_k,S_k$ by:
\begin{equation}
\label{rsRS}
  \begin{bmatrix} R_k+\Omega_k S_k \\ R_k-\Omega_k S_k \end{bmatrix}= \sqrt{\Omega_k} e^{-iv \tilde{x}} \begin{bmatrix} r_k +s_k \\ r_k^ \ast- s_k^\ast \end{bmatrix}\, .
 \end{equation}
With the differential operators
 \begin{equation}
 \label{diffop}
 \begin{split}
  \hat{Q} & \equiv \frac{1}{ \sqrt{2}} (\partial_x + 2 \kappa \tanh \kappa \tilde{x} )\\
  \hat{Q}^ \dagger  & \equiv \frac{1}{ \sqrt{2}} (- \partial_x + 2 \kappa \tanh \kappa \tilde{x})
\end{split}
 \end{equation} 
and the explicit form of the grey soliton background, the BdG equations \eqref{bdgmode} can be cast into the alternative form 
 \begin{equation}
 \label{altbdg}
\Omega_k \begin{bmatrix} r_k \\ s_k \end{bmatrix} =  \begin{bmatrix} i \sqrt{2} \beta  \hat{Q} &  \hat{Q}  \hat{Q}^ \dagger  +2 (\beta^2- \kappa^2) \\  \hat{Q}^ \dagger   \hat{Q} & -i \sqrt{2} \beta  \hat{Q}^ \dagger  \end{bmatrix} \begin{bmatrix} r_k \\ s_k \end{bmatrix} \; .
 \end{equation}
The displayed functions depend on $ \tilde{x} = x-x_0$. If we define a new function $ \Sigma$ by $  \hat{Q}^ \dagger  \Sigma \equiv s_k $, one obtains from the bottom line of \eqref{altbdg}
 \begin{equation}
 \label{firstsol}
 ( \Omega_k + i \sqrt{2} \beta  \hat{Q}^ \dagger ) \Sigma =  \hat{Q} r_k \; .
 \end{equation}
Acting with $\hat{Q}$ from the left on the top line of \eqref{altbdg} and using the relation \eqref{firstsol} and definitions \eqref{diffop}, the top line finally yields
\begin{equation}
\label{simbdg}
(\Omega_k^2 - i 2 \beta \Omega_k \partial_x - \frac{1}{4} \partial_x^4 + \kappa^2 \partial_x^2) \Sigma =0 \; .
 \end{equation}
Although this fourth order equation does not contain hyperbolic functions and looks very simple, it is equivalent to the coupled set of second order equations \eqref{altbdg}; we have not made any approximations. \eqref{simbdg} has the obvious solution $ \Sigma_k = C e^{i k x}$ with some normalization constant and the Bogoliubov frequency $ \Omega_k$ equal to \eqref{bogfre}. Applying the differential operator $\hat{Q}^ \dagger $ to $\Sigma_k$ directly gives us $s_k$, and lets us obtain $r_k$ by solving the first order equation \eqref{firstsol}. This verifies the non-zero frequency solutions \eqref{contsolr} and \eqref{contsols} of the preceding subsection.

\section{Zero frequency solutions of the BdGE}
\label{sec zero}
We now come to the zero frequency solutions, often referred to briefly as {\it zero modes}, which present special difficulties but are particularly important from some points of view. The zero frequency solutions solve the coupled system of equations
\begin{align}
\label{hom}
 0&= H_B R_ \eta + \psi_0^2 R_ \eta ^{ \ast} \\
\label{inh} m_\eta^{-1} R_ \eta &=  H_B S_ \eta - \psi_0^2 S_ \eta ^{ \ast}\\
\nonumber H_B  &= -\frac{1}{2} \partial_x^2 +i (\beta-v) \partial_x + 2 | \psi_0|^2 - \tilde{\mu} \; .
\end{align}
We will begin with the solutions for the infinite system without boundary conditions, and then derive the solutions for periodic boundary conditions. We will end this Section by discussing an interesting invariance property of the zero modes. 

\subsection{Infinite system without boundary conditions}
\label{sub inf}
A straightforward way to obtain solutions for the coupled set of equations \eqref{hom} and \eqref{inh} is to take the derivative of the GPE with respect to the five parameters $( \beta, v, x_0, \mu, \theta)$ of the background solution. All GPE solutions within this five-parameter family are time-independent, up to a uniform phase prefactor $e^{-i\tilde{\mu}t}$, and in some co-moving frame. Small perturbations within this family will therefore automatically have zero frequency. Those perturbations which do not shift $\tilde{\mu}$ or the lab-frame soliton velocity will be strictly time-independent. Those which do infinitesimally perturb $\tilde{\mu}$ will produce linear time dependence $\delta\psi \propto t$, since $e^{-i\delta\tilde\mu\, t}\doteq 1-i \delta\tilde\mu\,t$. Perturbations of the soliton's speed similarly induce a linearly time-dependent translation of the soliton, in any fixed frame. And linear time dependence is also zero frequency. 

These variations do not provide all zero-frequency BdGE solutions, but they do provide all the solutions needed for completeness. We can confirm this by finding the remaining linearly independent solutions, by other methods, and showing directly that they are unphysical. It is important to remember that for the infinite domain all five parameters are independent, apart from the condition $\kappa >0$; the periodic boundary condition that we subsequently consider will select a subspace out of this solution space, reducing the number of physical zero modes. 

We demonstrate the procedure with an example. Inserting the background solution \eqref{sol} into the GPE \eqref{GP1} and taking the derivative with respect to $ \theta$, yields
\begin{equation}
0 = \! \left( \!- \frac{1}{2} \partial_x^2 +i (\beta-v) \partial_x + 2 |\psi_0|^2 - \tilde{\mu} \! \right) \! i \psi_0 + \psi_0^2 (i \psi_0)^ \ast \; .
\end{equation}
Hence, we have found a first solution to the homogeneous equation \eqref{hom}:
\begin{equation}
\label{R1} R_1=  i \psi_0 \; .
\end{equation}

Taking the derivative with respect to the other solution parameters $x_0, \beta, v$ and $ \mu$, keeping in mind that we have abbreviated $ \tilde{\mu} = \mu +v \beta-v^2/2$, does not lead to all four solutions to \eqref{hom}, however. Varying $\tilde\mu$ or $\beta-v$ produces (linearly) time-dependent perturbations, unless we also change frame and the $e^{-i\tilde\mu t}$ prefactor in $\Psi$. Hence these perturbations do not yield further solutions to \eqref{hom}. Perturbations which do not affect $\tilde\mu$ or $\beta-v$, though, give two more solutions to the homogeneous equation:
\begin{align}
\label{R2} R_2 &= (i v - \partial_{x_0}) \psi_0 = (i v+ \partial_x) \psi_0 \\
\label{R3} R_3 & =-\partial_\beta \psi_0 -\partial_v \psi_0 +\beta \partial_\mu \psi_0\; .
\end{align}

Since \eqref{hom} is a system of two coupled second order linear differential equations, for the real and imaginary parts of $R$, there must exist a fourth linearly independent solution. We have obtained this by a variant of the factorization method used in subsection \ref{suzie} for the finite frequency modes (see Appendix \ref{derive zero}):
\begin{equation}
\begin{split}
\label{R4} R_4 = e^{-i v \tilde{x}}(& 3 \kappa \tilde{x} \sech^2 \! \kappa \tilde{x}+ 3 \tanh \kappa \tilde{x}+ \sinh 2 \kappa \tilde{x}\\
 &  +i \frac{4 \beta \kappa}{ \kappa^2 - \beta^2} \cosh^2 \! \kappa \tilde{x} ) \: ,
\end{split}
\end{equation}
with $\tilde{x}= x- x_0$. 

Turning now to the equation for the zero-modes' $S_{\eta}$, we can first identify four homogeneous solutions, which make the right-hand side of \eqref{inh} vanish. All four are immediately apparent: $S\to iR_{j}$. 

Then, for each of the four $R_{j}$ on the left side of  \eqref{inh}, there must exist corresponding particular solutions $S_{j}$ to the inhomogeneous equation.  Here we can make use of perturbations of $\tilde\mu$ and $\beta-v$, which generate $\delta\psi \propto t$.  Inserting the general zero mode time evolution
\begin{equation}
\begin{split}
\label{}
	p_\eta(t)&= p_\eta(0)\\
 q_\eta(t) &= q_\eta(0) + \frac{p_\eta(0)}{m_\eta}t
\end{split}
\end{equation}
into $\delta\psi = R_\eta q_\eta + i S_\eta p_\eta$, we can observe that perturbations to $\tilde\mu$ and $\beta-v$ must correspond to some combinations of zero-mode $p_{\eta}(0)$. It follows that derivatives of $\psi_{0}$ with respect to $\tilde\mu$ or $\beta-v$ will provide the components $S_\eta = -i\partial\delta\psi/\partial p_{\eta}$ of the zero mode $(R,S)$.  We are also free to include any combinations of the homogeneous solutions $iR_{j}$. Two simply expressed particular solutions that match $R_{1}$ and $R_{2}$ are respectively
\begin{align}
\label{S1}
S_1&=  i \partial_\mu \psi_0\\
\label{S2}S_2 &=i  \partial_\beta \psi_0 \; .
\end{align}
It can then simply be checked that $(R_{1},S_{1})$ and $(R_{2},S_{2})$ are indeed zero frequency solutions of the coupled BdGE \eqref{hom} and \eqref{inh}.

There are another two particular solutions to \eqref{inh}, with $R_3$ and $R_4$ on the left-hand side. In the appendix \ref{derive zero} we derive the asymptotic behavior for $x \to \pm \infty$ of these solutions $S_3$ and $S_4$, which is (as we will see) all we need for their consideration.

In this way we have found all eight linearly independent zero mode solutions, at least asymptotically. They are of the form $(R,S)\to(R_j,S_j)$ and $(R,S)\to(0,iR_j)$, for $j=1,2,3,4$. We can exclude most of these eight solutions as Hamiltonian modes, however. $(R_j,S_j)$ for $j=3,4$ and $(0,iR_4)$ are all unphysical, because one can prove from their asymptotic form that no linear combination of them can be normalizable according to \eqref{ortho} (they diverge exponentially at infinity). And we may discard without loss of generality the two zero frequency solutions $(0, iR_j)$ for $ j=1,2$, because their effect in $\delta\psi(x,t)$ could be absorbed by a shift of the canonical position variables $q_1$ and $q_2$. So, in case of the infinite system without boundary conditions, we are left with three possibly physical zero frequency solutions to the BdGE. 

Two of these remaining three solutions may be expressed as the linear combinations
\begin{equation}
\label{limitk}
\begin{bmatrix} R_{0^{ \pm}} \\ S_{0^{ \pm}} \end{bmatrix} \equiv \frac{1}{( 2 \pi \gamma_{\pm}c)^{\frac{1}{2}}}
\left( \gamma_{\pm} \begin{bmatrix} R_1 \\ S_1 \end{bmatrix}  \pm \frac{1}{2} \begin{bmatrix} R_2 \\ S_2 \end{bmatrix} \pm \begin{bmatrix} 0 \\ i R_3 \end{bmatrix} \right).
\end{equation}
Here $\gamma_{ \pm} \equiv c \mp \beta$ is the limit of $ \Omega_k/|k|$ for $k \rightarrow 0^{ \pm}$. One can show that \eqref{limitk} defines two positive mass zero frequency solutions which are $\delta$-normalizable. These two modes are precisely the limits $k\to0^{\pm}$ of the continuum of non-zero frequency solutions \eqref{contsolr} and \eqref{contsols}. Therefore they do not need to be separately added or considered; they are already fully represented within the continuum. 

In addition to these two positive mass modes we have a third linear independent mode, which we can take to be $(R_{2},S_{2})$. This has finite negative norm, and so we can identify
\begin{equation} 
\label{zero1}
 \begin{bmatrix} R_z \\ S_z \end{bmatrix} \equiv  \frac{1}{2 \sqrt{ \kappa}} \begin{bmatrix} -R_2 \\ S_2 \end{bmatrix}
\end{equation}
as a positive norm discrete mode with negative mass. This is the soliton translation mode, as one may see by recognizing that its effect in $\delta\psi$ is precisely to make a small translation of the soliton. 

We prove in Appendix \ref{completeness} that the continuum solutions alone are not complete, but that adding $(R_{z},S_{z})$ as one additional discrete mode provides completeness. We can therefore be confident that we have obtained all modes of the system as BdG solutions. There remains one point, though, which may seem to call for comment. Although we have identified three linearly independent zero frequency solutions, the three are not all orthogonal to each other according to our inner product as introduced in \eqref{ortho}. One source of this curious difficulty is the fact that $(R,S)\to(0,iR_{3})$ has zero norm according to the inner product. The reason that this subtle issue raises no problems for completeness is simply that the zero modes contained within the continuum are a set of zero measure. For any finite frequency, however small, there are two normalizable modes, and not three; and so the 'extra mode' $(0,iR_{3})$ at exactly zero frequency has no effect.

We attribute the appearance of the extra zero-norm solution to the fact that the boundary condition, which is only that $R$ and $S$ should not diverge too fast at infinity, is somehow too weak. If we replace the strictly infinite line with a periodic ring, however large, then the $(0,iR_{3})$ solution ceases to be independent of the other two, but instead must be added to them in particular proportions, in order to meet the periodic boundary condition. 

\subsection{Periodic boundary conditions}
\label{sub period}
Before deriving the periodic zero frequency solutions by taking derivatives of the GPE as in the preceding subsection, we want to emphasize two details. Firstly, we remind the reader that the results for the periodic system presented in the following are valid to order $ \mathcal{O}(e^{-2 \kappa L})$, which is an excellent approximation for $\kappa L \gg 1$. Secondly, the gas velocity $v$, the relative soliton speed $\beta$ and the chemical potential $\mu$ are not independent parameters for the periodic system, but must together satisfy a periodicity condition, as explained in Section \ref{greysol}. It turns out that cumbsersome expressions simplify most easily if we choose the relative soliton speed $ \beta$, and the chemical potential $\mu$ to be the two independent variables, in terms of which $v$ may be expressed. Hence, derivatives with respect to $\beta$ and $\mu$ are now different from the infinite case, because $v$ must now also be varied whenever $\beta$ and $\mu$ are.

 Inserting the grey soliton background \eqref{sol} into the GPE \eqref{GP1}, and differentiating this equation with respect to the remaining four independent parameters $( \theta, x_0, \beta, \mu)$ yields the four constituent functions $\tilde{R}_{1,2}, \tilde{S}_{1,2}$ of two solutions $(R,S)\to(\tilde{R}_{1,2},\tilde{S}_{1,2})$. Since the background solutions within this four parameter family are all periodic, all their derivatives with respect to these four parameters are automatically periodic as well. A linearly independent basis for these two periodic zero modes can be expressed as two different periodic linear combinations of the three non-periodic solutions for the infinite domain considered in the previous subsection \ref{sub inf}:
\begin{align}
\label{zerophas} \begin{bmatrix} \tilde{R}_{1} \\ \tilde{S}_{1} \end{bmatrix} &= \begin{bmatrix} R_1 \\ S_1 \end{bmatrix} - (\partial_\mu v) \begin{bmatrix} R_2-v R_1\\ S_2-\beta S_1+i R_3 \end{bmatrix}  \\
\label{zerotransp} \begin{bmatrix} \tilde{R}_{2} \\ \tilde{S}_{2} \end{bmatrix} &= \begin{bmatrix} R_2 \\ S_2 \end{bmatrix} -(\partial_\beta v) \begin{bmatrix} R_2-v R_1\\ S_2-\beta S_1+i R_3 \end{bmatrix} \; .
\end{align}
The explicit expressions for $\partial_\mu v $ and $\partial_\beta v $ can be found by differentiating \eqref{velocity} with respect to the corresponding parameters and are both of order $ \mathcal{O}(1/(\kappa L))$:
\begin{align}
\partial_\mu v &= -\frac{\beta}{ 2 \kappa L \mu}\; , & \partial_\beta v &= \frac{1}{ \kappa L} \; .
\end{align}

Besides these two periodic zero frequency solutions $(\tilde{R}_{1,2}, \tilde{S}_{1,2})$, and the corresponding homogeneous solutions $(0, i \tilde{R}_{1,2})$ that can be absorbed into shifts of $q_{1,2}$, there must exist four more zero frequency solutions. Since in comparison with subsection \ref{sub inf} above we have changed only the boundary conditions and not the equation, these solutions must simply be $(R_{3,4},S_{3,4})$ and $(0,iR_{3,4})$, where $(R_{j}, S_{j})$ are given in Appendix \ref{derive zero}.  It is not difficult to show that no combinations of these four solutions can be made smoothly periodic, and so $(R_{1,2},S_{1,2})$ span all the zero modes that we need to canonically represent $\delta\psi(x,t)$.

It remains only to select combinations of these solutions, and possibly perform the mass-changing transformation $(R_{j},S_{j})\to (-R_{j},S_{j})$, to assemble from our two-solution basis a pair of positive norm solutions that are orthogonal to each other.  This is accomplished with
\begin{align}
\label{leadphas} \begin{bmatrix} \tilde{R}_{0} \\ \tilde{S}_{0} \end{bmatrix}  & \equiv \frac{i}{ \sqrt{2 L}} \begin{bmatrix}  R_1 \\ S_1 \end{bmatrix} + \mathcal{O} \left( \frac{1}{(\kappa L)^{3/2}} \right ) \\
\label{leadtrans} \begin{bmatrix} \tilde{R}_{z} \\ \tilde{S}_{z} \end{bmatrix}  
& \equiv \frac{1}{2 \sqrt{ \kappa}} \begin{bmatrix} -R_2\\ S_2\end{bmatrix} + \mathcal{O} \left( \frac{1}{\kappa L} \right )\;.
\end{align}
Here the corrections of higher order in $1/( \kappa L)$ may be worked out explicitly, but we will assume that $L$ is large enough to ignore them. The negative mass solution \eqref{leadtrans} is equal to the corresponding solution \eqref{zero1} for the infinite system, up to the indicated order, so we use the same subscript $z$. It is the zero mode associated with the translation of the soliton. The periodic positive mass solution \eqref{leadphas}, labelled with subscript $0$, has been normalized and can be identified with the usual phase translation zero mode, which is present for any stationary Gross-Pitaevskii solution, to leading order in $1/( \kappa L)$. 

\subsection{Symmetry of the zero modes}
\label{sub sym}
The main result of this paper has now been achieved: we have presented a complete set of BdG solutions, including zero modes, for any grey soliton background. The only remaining question is whether our solution is unique. Since all our non-zero frequency modes are in degenerate pairs, it is obvious that arbitrary linear combinations of each degenerate pair may be taken; this trivial freedom is the only one remaining for the finite frequencies. 

The two zero modes of the finite system, namely the soliton translation and the phase translation mode, can also be mixed into each other, but the transformation is a bit different. It is straightforward to show that the BdGE \eqref{hom} and \eqref{inh} are also solved by the following new zero-frequency solutions for any real $\theta$:
\begin{equation}
\label{sinhrot}
\begin{bmatrix} R_0 ^{ \prime} \\ R_z^ { \prime} \\ S_0 ^{ \prime} \\ S_z ^{ \prime} \end{bmatrix} = \begin{bmatrix}  \cosh \theta  & -\sinh \theta & 0 & 0 \\ - \sinh \theta & \cosh \theta & 0 & 0  \\  0 & 0 & \cosh \theta & \sinh \theta \\ 0 & 0 & \sinh \theta  & \cosh \theta \end{bmatrix}  \begin{bmatrix} R_0 \\ R_z \\ S_0 \\ S_z \end{bmatrix}
 \end{equation}
The two solutions are again of opposite mass. Considering \eqref{dev} and \eqref{hamfunction} one can see that this transformation leaves $ \delta \psi$ unchanged, and is equivalent to a proper canonical transformation of the $q_{1,2},p_{1,2}$ for the two zero frequency modes. 

While this symmetry is thus in one sense trivial, it is also surprising. It mixes the soliton motion and global phase translation degrees of freedom; and moreover these are modes of opposite mass. One may be tempted to suppose that positive and negative mass modes could never be confused, because they would have drastically different behavior. But in fact they may not always be so distinct, after all: the zero modes can be mixed by the canonical transformation associated with \eqref{sinhrot}.

\section{Discussion}
\label{discuss}
This has been a technical paper on a technical subject. Since there are not many exact but non-trivial BdG solutions known, we have tried to use our solution to provide a pedagogical discussion of important aspects of the general BdG problem. Many of the properties of our solutions, including the fact that we were able to find them, seem to depend on particular detailed properties of certain hyperbolic trigonometric functions, and in this sense are purely technical features of a special case. It is in particular disappointing to report that, although the grey soliton GPE solution itself may be extended into a class of multi-soliton solutions given exactly by certain elliptic functions, and this class may be shown to include all time-independent GPE solutions with uniform external potential in one dimension, we have been unable to extend our exact BdG solutions to these cases. The solutions we have found seem to depend crucially on properties of $\tanh$ and $\sech^2$ that are not shared by their corresponding elliptical generalizations. But the problem of BdG normal modes in a grey soliton background is interesting from some rather fundamental viewpoints, and we will close our paper by discussing two of them.

A feature of the stationary grey soliton solutions which has on occasion caused excitement is the fact that they seem to include sonic event horizons.  The local speed of sound in the hydrodynamic approximation to the GPE is $|\psi|$; the local fluid speed is $\partial_{x}\arg{\psi}$. It is easy to show that this local fluid speed exceeds this local sound speed within a finite range $|x-x_{0}|<x_{h}$ for a certain $\beta$-dependent $x_{h}$, apparently providing a black + white hole scenario with horizons at $x-x_{0}=\pm x_{h}$. The very problem we have solved would therefore seem to be an ideal candidate for terrestrial investigations of the Hawking effect in black holes, by examining quasiparticle pair production after quantization. Unfortunately, however, our exact results show that this does not work: grey solitons do not really have horizons. No perturbation wavepackets are reflected at any point in the soliton background, regardless of which direction they move in; modes of any wavelength can freely pass through the supposed horizons, in any direction. 

This disappointingly ordinary behavior illustrates an important caveat which must never be overlooked in constructing black hole analogs in fluids. Sound waves only imitate light, with a maximum speed and no dispersion, in the hydrodynamic regime. This means that the wavelength must not be too short, so that the hydrodynamic picture does not break down. Yet even for real black holes and real light, the horizon concept is only applicable for wavelengths short in comparison with the length scale over which the background changes, so that the geometric optics limit in which light propagates on null geodesic rays becomes valid. Light with a wavelength much longer than a real black hole's Schwarzschild radius is not trapped by the event horizon, but simply diffracted. The requirement for a sonic horizon is therefore that there exist a range of wavelengths which are simultaneously long enough to be hydrodynamic, but short enough to be geometric. For a grey soliton it turns out that there are no such wavelengths: perturbations are all either non-hydrodynamic or non-geometric. Our exact solutions then merely confirm the fears of horizonless behavior which will be raised by careful estimates of validity regimes.

Another fundamental question which grey soliton BdG solutions can address is the emergence of collective co-ordinates in classical fields and quantum many-body systems \cite{Anglin2008}. Such collective co-ordinates represent many fundamentally interesting and even practically important degrees of freedom in the real world. The mental degrees of freedom of conscious beings, for example, are presumed by substance monists to be collective co-ordinates of the large and intricate quantum many-body systems known as brains. Such co-ordinates stand out in some way from the quasi-continuum of other degrees of freedom, but they do not entirely decouple from them; the closest we can accurately come to ignoring all those less interesting other modes is to treat them as an environment or reservoir, and describe the evolution of the collective co-ordinates as a kind of generalized Brownian motion. This is understood to introduce thermodynamics and quantum decoherence, but understanding exactly how this happens and what it means is a goal still unachieved.

Perhaps the most basic question is simply, What is it that makes the collective co-ordinates stand out from the quasi-continuum crowd? One might call this very basic issue `the Hamiltonian \textit{gavagai} problem', after the philosopher W.V.O. Quine's identification of a similar problem in linguistics \cite{Quine1960}. Quine imagined that a linguist learning an unknown language from a native speaker might see a rabbit run across a field, and hear the native exclaim, ``Gavagai!'' While the hypothesis that `gavagai' was simply their language's word for `rabbit' might seem obvious, Quine pointed out that it might just as well mean some part of a rabbit, or some aspect of the rabbit's motion, or some relationship between the rabbit and the field, or any number of other concepts. While Quine's `gavagai problem' was formulated as a problem in the theory of translation between languages, it has become a touchstone in the philosophy of meaning. A much simplified but basically similar problem presents itself in the Hamiltonian dynamics of many degrees of freedom. Infinitely many canonical coordinate systems are possible; some are evidently better than others; but we lack an explicit theory to determine which ones might be best. In particular, although it is a familiar fact that a small number of degrees of freedom often stand out as prominent, we lack a theory of prominence. 

While it is possible that such problems simply lie outside the subject of physics, and will remain grist for philosophy, they do seem difficult to avoid when discussing important physical processes, such as dissipation and decoherence, at a fundamental rather than phenomenological level. Perhaps it is worth looking for a dynamical theory of prominence which might, like adiabatic theory and in the study of dynamical chaos, prove both rigorous and rich, despite being based on concepts that were not originally seen as part of mechanics.

Such a goal surely lies well beyond the kind of linearized analysis we have presented in this paper, since linear modes can always be exactly decoupled. Within linear theory, there is no obvious reason why a collective co-ordinate should not simply be one of the normal modes, and hence decouple exactly from all the other modes, after all. Yet the linear regime may still offer some insights and hints. 

In the first place we confirm that the problem is subtle. The motion of the soliton, which is the obvious collective degree of freedom in this problem, is indeed represented as one of the BdG normal modes, namely the negative mass zero mode. But we have seen that there is no absolute distinction between this mode and the other zero mode, which is a long wavelength background phase gradient mode that is present even in the absence of the soliton. A canonical transformation can mix the two modes, while preserving exactly the form of their linearized Hamiltonian. One can conjecture that nonlinear dynamics, which in the GPE is local in space, may somehow prefer that canonical representation of the two zero modes which makes the soliton motional mode as spatially localized as possible. Since the soliton's phase pattern is very nonlocal, however, even though its modulus deformation is localized exponentially well, the most localized zero mode possible is still not perfectly localized, but does include long range perturbations of the phase of $\psi$. Yet since those perturbations are spatially uniform except near the soliton, and a uniform phase shift is in most contexts unobservable, it is not clear to what extent this represents a `genuine' long range extension, rather than something like a topologically nontrivial gauge configuration.

It is instructive to contrast the subtlety of these issues for the grey soliton background with their triviality in the bright soliton case, where the sign of the $|\Psi|^{4}$ term in \eqref{hamilton0} is negative rather than positive. A bright soliton is a localized blob with $\psi\sim\sech\kappa(x-x_{0})$. Rather than moving through an asymptotically uniform background field, it is itself the entire field; far from $x_{0}$, the field vanishes exponentially. The bright soliton can move, but its motion is identical to the motion of the entire configuration; there is no possibility of relative motion between the soliton and the background, as there is the grey soliton case. Dynamically, this difference shows up in the fact that the grey soliton's motional zero mode lies at the bottom of a gapless continuous spectrum of other modes, while the bright soliton's zero mode is separated in frequency from the continuum by a finite gap of width $| \mu|$. (See Appendix \ref{brightbog}, below.) Distinguishing soliton from background, as a collective degree of freedom, can therefore be done easily and unambiguously for the bright soliton, because of this large separation in time scales. For the grey soliton there is no time scale separation, but only a contrast in the pairing of time and length scales. The grey soliton moves much more slowly than any other perturbations with wavelengths on the order of its size, and it is much smaller in extent than any other perturbations that evolve as slowly as it moves. There is therefore reason to hope that the role of locality in nonlinear GPE dynamics will indeed single out a unique soliton collective degree of freedom; but there is no obvious identification based on time scales within the linear regime.

The technical difficulties in going beyond the linear approximations used in this paper are considerable, but they are joined by conceptual difficulties as well, particularly in the quantum problem. The difficult issue is backreaction: since the continuum modes do interact with collective degrees of freedom, beyond the linear regime, we must expect that the collective co-ordinate evolution is in some way `dressed' or perturbed by the other modes. Yet if these modes are quantum mechanical, or even if they are classical but subject to thermal fluctuations, then one seems to be faced with the challenge of somehow incorporating thermal or quantum fluctuations into the collective co-ordinate itself. This may be feasible, though it might imply large fluctuations or even a `Schr\"odinger's Cat' state; but it is not clear whether it is really correct, or whether perhaps instead the right definition of the collective degree of freedom must in some way average over these fluctuations, in the same way that a temperature or a center of mass involves an averaging by definition. This issue is not purely philosophical, but may in principle have clearly observable consequences. In a series of single observations of an identically prepared quantum or thermal system containing a soliton, do we expect to see in each run a smeared out soliton in the same location? Or do we expect to see a sharply resolved soliton, whose location fluctuates randomly from run to run?

Nonlinear perturbation dynamics in grey soliton backgrounds may thus provide a simple model for issues of fundamental importance. By providing a complete and explicit analytic solution of the linearized problem in this case, we have laid a foundation on which that further study can be built.


\begin{appendix}

\section{Orthonormality}
\label{orthonormality}
We will first show orthonormality for the infinite system; the finite periodic case is straightforwardly similar. If we define for complex functions $f(x)$, $g(x)$ 
\begin{equation}
\langle f|g \rangle \equiv \int \! \! dx \left( f g^{ \ast} + f^{ \ast} g\right)
\end{equation}
with integration over the infinite or finite domain, respectively, then the orthonormality relations \eqref{ortho} we have to confirm are equivalent to
\begin{align}
\label{rs} \langle R_\eta | S_\xi \rangle &= \delta_{ \eta \xi}\\
\label{rr} \langle R_\eta | i R_\xi \rangle &= 0\\
\label{ss} \langle S_\eta | i S_\xi \rangle& = 0\;.
\end{align} 
The Kronecker delta is to be understood as a Dirac delta function $\delta(k-k')$ whenever $\eta\to k$ is continuous (in which case all three integrals are distributions, rather than true functions of $k,k'$). Where the discrete mode is involved, we intend $\delta_{zz}=1$, and $\delta_{kz}=\delta_{zk}=0$ (also as a distribution in $k$). 

On the one hand we have directly evaluated these integrals, from our explicit expressions for $R_{\eta}, S_{\eta}$ and could prove the orthonormality relations \eqref{ortho} for continuous index $k$ and the discrete zero mode with index $z$ in a lengthy but straightforward calculation. It is much easier, however, to prove orthonormality for modes with $ \Omega_\eta^2 \neq \Omega_\xi^2$ using the BdGE \eqref{bdgmode} and integration by parts. We demonstrate this for \eqref{rr}:
\begin{equation}
\label{exprrss}
\begin{split}
i \langle R_\eta | iR_\xi \rangle &= m_\xi \int \! \! dx \left[ R_ \eta (H_B S_\xi- \psi_0^2 S_\xi^{ \ast})^ \ast \right. \\ & \qquad \qquad \quad \left. - R_\eta^\ast (H_B S_\xi- \psi_0^2 S_\xi^{ \ast}) \right]\\
&= m_\xi \int \! \!  dx \left[ S_\xi^\ast (H_B R_\eta+ \psi_0^2 R_\eta^{ \ast}) \right. \\ & \quad \qquad \qquad \left. - S_\xi (H_B R_\eta+ \psi_0^2 R_\eta^{ \ast})^ \ast \right] \; .
\end{split}
\end{equation}
In the first line we have used the BdGE \eqref{bdgmode} for $R_\xi$ and in the second line we have integrated by parts. Once again using the BdGE yields
\begin{equation}
\label{rrss}
i \langle R_\eta | iR_\xi \rangle = m_\xi m_\eta \Omega_\eta^2 \langle S_\eta| i S_\xi \rangle \; .
\end{equation}
As a first result we find that for non-zero frequency modes the relations \eqref{rr} and \eqref{ss} are not independent of each other. Redoing the procedure \eqref{exprrss} for \eqref{rr}, but now using the BdGE for $R_\eta$ instead, reveals
\begin{equation}
i \langle R_\eta | iR_\xi \rangle = m_\xi m_\eta \Omega_\xi^2 \langle S_\eta| i S_\xi \rangle \; .
\end{equation}
Hence, for $\Omega_\xi^2 \neq \Omega_\eta^2$, one can deduce \eqref{rr}. With this result also \eqref{ss} is proven by \eqref{rrss}.

Along similar lines one can confirm \eqref{rs}. In a first step it is easy to show that
\begin{align}
\label{switching}
\langle R_\eta | S_\xi \rangle &= \frac{ m_{ \eta}}{m_{\xi}} \langle S_\xi | R_\eta \rangle \; .
\end{align}
It follows that the orthonormality relation \eqref{rs} is symmetric under switching $ \eta \leftrightarrow \xi$ up to the constant factor. In analogy to the above proof it is a straightforward exercise to confirm \eqref{rs}.

The results found for the infinite system can be transferred to the finite system with periodic boundary conditions in the following way. The normalization constant $ \mathrm{N}_k$ has to be changed (see \eqref{defnorm}) and the delta-distribution in \eqref{ortho} turned into a Kronecker-delta, while the allowed non-zero values of the index $ k$ are given by the quantization condition \eqref{conditionk}. The two zero frequency modes presented in subsection \ref{sub period} can easily be made orthonormal in a Gram-Schmidt procedure, and they are naturally orthogonal to all modes with non-zero frequency.

\section{Completeness}
\label{completeness}
For the infinite system without boundary conditions, the completeness relations \eqref{completerel} for the set of continuous Bogoliubov modes \eqref{littler}, \eqref{littles} and the discrete zero mode \eqref{zero1} can be proven directly, by straightforward contour integrations over the continuous index $k$. We express the continuous Bogoliubov modes in terms of the functions $r_k$ and $s_k$ given in \eqref{contsolr} and \eqref{contsols}. Both relations then involve terms of the form
\begin{equation}
\label{genericcomplete}
\begin{split}
\int \! \! dk \left[ r_k (x) s_k ^{ \ast} (y) + r_k  ^{ \ast} (x) s_k  (y) \right] &= \delta (x - y) -\frac{ \kappa^2}{2} \sech^2 \! \kappa x\\
\int \! \! dk  \left[ s_k  (x) s_k  ^{ \ast} (y) - s_k  ^{ \ast} (x) s_k  (y) \right] &=0\\
\int \! \! dk  \left[ r_k  (x) r_k  ^{ \ast} (y) - r_k  ^{ \ast} (x) r_k  (y) \right] &\\
& \hspace{-3cm} = -\frac{ i \beta}{2} \left(\tanh \kappa x \sech^2 \! \kappa y - \tanh \kappa y \sech^2 \! \kappa x \right )\\
& \hspace{-2.5cm} + \kappa (x- y )\sech^2 \! \kappa y \sech^2 \! \kappa x \; .
\end{split}
\end{equation}
Simplifying the results of the contour integrations to reveal the above simple expressions requires some judicious application of identities among hyperbolic trigonometric functions, which the authors have successfully performed, but the exercise is involved and we omit the details here. Numerical plotting of the unsimplified results will quickly confirm \eqref{genericcomplete}.

The fact that the right-hand sides of \eqref{genericcomplete} is not of the simple form required by canonicity and completeness proves that the continuous BdG modes are not complete by themselves. But it easy to see that this defect is remedied precisely by adding one more mode, namely the discrete negative mass zero mode associated with soliton motion, $(R_{z},S_{z})$. Thus we have proven that the discrete negative mass zero mode and the continuous modes form a complete set of functions. 

For the finite system with boundary conditions the integral over the modes becomes a sum over the allowed indices (cf.~Eq.~\eqref{conditionk}) plus the periodic negative and positive mass zero mode we have found in \ref{sub period}. A direct proof that the summation gives a partition of unity is not accessible due to the complicated quantization condition \eqref{conditionk}. But it seems plausible that we can decompose any function that satisfies periodicity on the ring in terms of the periodic functions of our complete set.

\section{Orthonormality from completeness}
\label{proof}
Here we show that the orthonormality relations \eqref{ortho} are a consequence of completeness and canonicity \eqref{completerel}, together with uniqueness, i.e.~invertibility of the mapping from $\{q_{\eta},p_{\eta}\}$ to $\delta\psi(x)$. 

We begin by defining the generalized matrix
\begin{equation}
\label{}
	A_{x,\eta} \equiv \begin{bmatrix} R_\eta (x) & i S_\eta (x) \\ R^*_\eta (x) & -i S^*_\eta(x)\end{bmatrix}\;.
\end{equation}
Here $\eta$ is a Bogoliubov mode index; summing over $\eta$ is to mean summing over all discrete Bogoliubov modes $k,z$ in the periodic case, or integrating over continuous $k$ and then adding the discrete zero mode $\eta=z$ in the infinite version of the problem. We call $A_{x,\eta}$ a generalized matrix simply because we want to consider $x$ and $\eta$ as two matrix indices, even though $x$, and possibly $\eta$ as well, is continuous. In a similar sense we define also the generalized matrix
\begin{equation}
\label{}
	B_{\eta,x} \equiv \begin{bmatrix} S^*_\eta (x) & S_\eta (x) \\ -iR^*_\eta (x) & i R^*_\eta(x)\end{bmatrix}\;.
\end{equation}
These two matrices $A$ and $B$ are constructed so that the `matrix product' $AB$ gives the completeness relation:
\begin{equation}
\label{compNAB}
	N_{x,y}\equiv \sum_\eta A_{x,\eta}B_{\eta,y} = \begin{bmatrix} \delta(x-y) & 0 \\ 0 & \delta(x-y)\end{bmatrix}\;,
\end{equation}
where as usual we mean the sum over $\eta$ to be interpreted as an integral in case $\eta$ is continuous.

Orthonormality, on the other hand, concerns the matrix product $BA$:
\begin{equation}
\label{orthoMAB}
	M_{\eta,\xi}\equiv \int\! dx\, B_{\eta,x} A_{x,\xi}\;.
\end{equation}
Simply working out the matrix product here shows that the Bogoliubov orthonormality property, which we are trying to demonstrate, is just that $M$ is in fact the identity matrix. This we now prove, in two steps.

First we show that $M$ is a projection operator, and so all of its eigenvalues must be either 0 or 1.
To see this we simply compute
\begin{equation}
\begin{split}
	[M^2]_{\eta,\xi}&= \sum_\lambda M_{\eta,\lambda} M_{\lambda,\xi} \\
	&= \int\! dx\,dy\, \sum_\lambda\, B_{\eta,x} A_{x,\lambda} B_{\lambda,y}A_{y,\xi} \\
	&=  \int\! dx\,dy\, B_{\eta,x} N_{x,y} A_{y,\xi} \\
	&= \int\!dx\, B_{\eta,x} A_{x,\xi} = M_{\eta,\xi}\;.
\end{split}
\end{equation}
To go from the third to the fourth line here we use the completeness relation \eqref{compNAB}. We have therefore shown that $M^2 = M$, which implies that the eigenvalues of $M$ can only be zero or one.

Our second step is to show that $M$ is invertible, and that it can therefore have no zero eigenvalues. We do this by invoking the assumption of invertibility of the mapping from the $\{q_\eta, p_\eta\}$ to $\psi(x)$, which implies that there exists some generalized matrix $C_{\eta,x}$, of the same form as $B_{\eta,x}$, such that
\begin{equation}
\label{}
	\begin{bmatrix} q_\eta \\ p_\eta\end{bmatrix} = \int\!dx\, C_{\eta,x} \begin{bmatrix} \delta\psi(x) \\ \delta\psi^*(x) \end{bmatrix}\;. 
\end{equation}
We then invoke our basic expression for $\delta\psi(x)$ as a function of the $\{q_\eta, p_\eta\}$, Eqn.~\eqref{dev}, which may be compactly expressed using our generalized matrix $A$:
\begin{equation}
\label{}
	\begin{bmatrix} \delta\psi(x) \\ \delta\psi^*(x) \end{bmatrix} = \sum_\eta A_{x,\eta}\begin{bmatrix} q_\eta \\ p_\eta\end{bmatrix}\;. 
\end{equation}
Combining these two expressions yields
\begin{equation}
\begin{split}
	\begin{bmatrix} q_\eta \\ p_\eta\end{bmatrix} &= \sum_\xi \int\!dx\, C_{\eta,x} A_{x,\xi}\begin{bmatrix} q_\xi \\ p_\xi\end{bmatrix} \\
	&=\sum_\xi  \int\!dx\,dy\, C_{\eta,x} N_{x,y} A_{y,\xi}\begin{bmatrix} q_\xi \\ p_\xi\end{bmatrix} \\
	&\equiv \sum_{\xi,\lambda} L_{\eta,\lambda} M_{\lambda,\xi} \begin{bmatrix} q_\xi \\ p_\xi\end{bmatrix}\;. 
\end{split}
\end{equation}
where we have again used completeness \eqref{compNAB} to go from the first line to the second, and in the last line we simply define
\begin{equation}
	L_{\eta,\xi} \equiv \int\!dx\, C_{\eta,x} A_{x,\xi}\;.
\end{equation}
This final result holds for all possible $q_\eta, p_\eta$, which implies that $LM$ is the identity matrix. Hence the inverse of $M$ exists (because we have constructed it, from the assumed $C$, as $L$).

Since all eigenvalues of $M$ are therefore either zero or one, and since none of them can in fact be zero, it follows that they are all one. Thus $M$ is the identity matrix, which is the orthonormality condition. This further implies that $C=B$.

\section{Derivation of the zero frequency solutions}
\label{derive zero}
We present here in very detail how to obtain all (asymptotic) zero frequency solutions of the BdGE \eqref{hom} and \eqref{inh}. This is best achieved by rewriting the BdGE analogous to the factorization method of subsection \ref{suzie}. Without loss of generality we set the soliton position $x_0$ to zero in this appendix.

At first we solve the homogeneous equation \eqref{hom} with the ansatz $R(x)= e^{-iv x} (f(x)+ i g(x))$ with real functions $f$ and $g$. With the definition of the operators $\hat{Q}$ and $ \hat{Q}^ \dagger $ from \eqref{diffop}, the second order differential equation in complex function space can be rewritten as a coupled system in real function space:
\begin{equation}
\label{Rmatrix}
\begin{bmatrix} 0 \\ 0 \end{bmatrix} = \begin{bmatrix} \sqrt{2} \beta \hat{Q}& 2( \beta^2- \kappa^2) +\hat{Q} \hat{Q}^ \dagger \\ \hat{Q}^ \dagger  \hat{Q} & \sqrt{2} \beta \hat{Q}^ \dagger  \end{bmatrix} \begin{bmatrix} f \\ g \end{bmatrix}\; .
\end{equation}
Except for the imaginary unit the matrix involved is identical to the one in \eqref{altbdg}. From the bottom line one finds that 
\begin{equation}
\label{Qf}
\hat{Q} f= - \sqrt{2} \beta g+\frac{C_4 \kappa}{\sqrt{2}} \cosh^2 \! \kappa x \; ,
\end{equation}
where the second term on the right hand side is annihilated by $ \hat{Q}^ \dagger $ for arbitrary constant $ C_4$. Inserting this in the top line of \eqref{Rmatrix} it can be cast into an inhomogeneous second order differential equation for $g$:
\begin{equation}
\label{geq}
 \left(-\frac{1}{2} \partial_x^2 - \kappa^2 \sech^2 \! \kappa x \right) g = -\kappa \beta C_4 \cosh^2 \! \kappa x \; .
\end{equation}
 If we define 
\begin{equation}
\begin{split}
 g_1 &\equiv  \tanh \kappa x \qquad \qquad  g_3  \equiv \kappa x \tanh \kappa x-1 \\
 g_4 & \equiv \frac{\beta}{\kappa} \cosh^2 \! \kappa x
\end{split}
\end{equation}
one finds that $g_1$, $g_3$ are the complementary functions of the differential equation \eqref{geq} and $g_4$ is the particular solution. Hence, the general solution is given by
\begin{equation}
\label{generalg}
g = \sum_{j=1 \atop j \neq 2}^4 C_j g_j \; ,
\end{equation}
where $C_1$, and $C_3$ are arbitrary constants and $C_4$ is the same solution parameter as in \eqref{Qf}.

With the general solution \eqref{generalg} one can search for a solution $f$ in \eqref{Qf}. The general solution to this inhomogeneous first order equation is straightforwardly found to be
\begin{equation}
\label{generalf}
f = \sum_{j=1}^4 C_j f_j \; ,
\end{equation}
where we have used the definitions 
\begin{equation}
\label{ff}
\begin{split}
 f_1 & \equiv - \frac{ \beta}{\kappa} \hspace{2cm} f_2 \equiv \sech^2 \! \kappa x  \\
 f_3 &\equiv \frac{ \beta}{ \kappa}  \left( \frac{3}{2} \left( \tanh \kappa x + \kappa x \sech^2 \! \kappa x \right)-\kappa x \right) \\
 f_4 &\equiv \frac{\kappa^2-\beta^2}{4 \kappa^2} \left(\sinh 2 \kappa x+ 3 (\tanh \kappa x + \kappa x \sech^2 \! \kappa x)\right)
 \end{split}
 \end{equation}
and introduced $C_2$ as solution coefficient for the complementary function $f_2$. This yields the four complementary solutions $R_j=e^{-iv x}(f_j+i g_j)$ (with $g_2 \equiv 0$) for the homogeneous BdGE \eqref{hom} as presented in subsection \ref{sub inf}.

In the same manner one can transform the inhomogeneous BdGE \eqref{inh} with the ansatz $ S \equiv e^{-iv x}(i q-p)$, where $q$ and $p$ are real functions, into 
\begin{equation}
\label{Smatrix}
\begin{bmatrix} -f \\ g \end{bmatrix} = \begin{bmatrix} \sqrt{2} \beta \hat{Q}& \hat{Q} \hat{Q}^ \dagger +2( \beta^2- \kappa^2) \\ \hat{Q}^ \dagger  \hat{Q} & \sqrt{2} \beta \hat{Q}^ \dagger  \end{bmatrix} \begin{bmatrix} q \\ p \end{bmatrix} \; ,
\end{equation}
with the inhomogeneities given by \eqref{generalg} and \eqref{generalf}. The lower line can be viewed as an inhomogeneous first order differential equation for the function $\hat{Q} q+ \sqrt{2} \beta p$. Its particular solution can be shown to be 
\begin{equation}
\label{Qq}
\begin{split}
\sqrt{2} \hat{Q} q+ 2 \beta p & = \frac{C_1}{ \kappa} + \frac{C_3}{ \kappa} ( \kappa x + \frac{1}{2} \sinh 2 \kappa x)\\
& \hspace{0.5cm} - C_4 \frac{2 \beta}{ \kappa} x \cosh^2 \! \kappa x \; ,
\end{split}
\end{equation}
with the same constants $C_j$ as in \eqref{generalg} and \eqref{generalf}. The homogeneous part of this equation is solved by $g_4$, which would finally lead to a solution $(0,iR_4)$ of the complete BdGE \eqref{hom} and \eqref{inh}. As we already know from \ref{sub inf} that the solutions of the type $(0, iR_j)$ exist, we do not consider it in the following, but concentrate on finding the particular solutions.

Inserting \eqref{Qq} into the upper line of \eqref{Smatrix} yields a second order differential equation for $p$ of the form \eqref{geq}, but with different inhomogeneities:
\begin{equation}
\label{inhp}
\begin{split}
& \left(-\frac{1}{2} \partial_x^2 - \kappa^2 \sech^2 \! \kappa x \right) p\\
=&-C_2 f_2-C_3 \left(f_3+\frac{\beta}{ \kappa} \left( \kappa x + \frac{1}{2} \sinh 2 \kappa x \right) \right)\\
&-C_4 \left(f_4- \frac{2 \beta^2}{ \kappa}x \cosh^2 \! \kappa x \right) \; .
\end{split}
\end{equation}
 Thus the complementary functions for the homogeneous part of this equation are exactly the same functions $g_1$ and $g_3$, and the remaining problem is to determine the particular solutions for $p$. We decompose $p = \sum_{j=1}^4 C_j p_j$, where the coefficients $C_j$ are chosen to be the same as in \eqref{generalg} and \eqref{generalf}, and solve the equation \eqref{inhp} for each individual $j$. Since \eqref{inhp} does not contain an inhomogeneity with $C_1$, the corresponding solution $p_1$ is zero, and for $j=2$ one easily finds the particular solutions $p_2$, such that we have:
\begin{align}
p_1 & = 0 \; , & p_2&= \frac{1}{ \kappa^2 } \; .
\end{align}
We have not found the exact solutions for $p_3$ and $p_4$. Nevertheless, one can expand \eqref{inhp} in powers of $e^{2 \kappa |x|}$ to determine their asymptotic behavior for $ x \to \pm \infty$. This will be sufficient enough for our purpose since one can prove from their asymptotic form that they do not lead to physical zero modes, i.e.~these modes are either not properly normalizable in the case of the infinite system without boundary conditions or they are incompatible with periodicity in case of the finite system. The leading terms are given by
\begin{align}
\label{p3} p_3 & \simeq \frac{\beta}{8 \kappa^3} \sgn x e^{2 \kappa |x|} \\
\label{p4} p_4 & \simeq \left( \frac{\kappa^2 + 3 \beta^2}{16 \kappa^4} \sgn x- \frac{ \beta^2}{4 \kappa^3} x \right) e^{2 \kappa |x|} 
\end{align}
and are due to the exponentially growing inhomogeneities in \eqref{inhp}.

With the given (asymptotic) solutions $p_j$ one can solve the first order equation \eqref{Qq} in a straightforward manner with the ansatz $q= \sum_{j=1}^4 C_j q_j$. The particular solutions for $j=1,2$ are exactly given by
\begin{align}
 q_1 &= \frac{1}{2 \kappa^2} \left( \tanh \kappa x + \kappa x \sech^2 \! \kappa x \right)=-\frac{ \kappa }{2\beta} q_2 
\end{align}
and for $j=3,4$ we find for $ x \to \pm \infty$
\begin{align}
\label{q3} q_3 & \simeq \frac{\kappa^2-\beta^2}{16 \kappa^4} e^{2 \kappa |x|} \\
\label{q4} q_4 & \simeq \frac{\beta}{8 \kappa^4} \left( \left(\beta^2-\kappa^2 \right) x \sgn x - \frac{ \beta^2}{\kappa} \right) e^{2 \kappa |x|} 
\end{align}
to leading order in $e^{2 \kappa |x|}$. Combining the results to $S_j =e^{-iv x}( iq_j-p_j)$ yields the exact zero frequency solutions to \eqref{inh} with inhomogeneities $R_j$ for $j=1,2$, which are equivalent to the solutions presented in \eqref{S1} and \eqref{S2}. In the case $j=3,4$ we correspondingly find the asymptotic solutions. It is important to recognize that we have found the asymptotic behavior of global solutions $S_3$ and $S_4$.


\section{Linearization for a bright soliton background}
\label{brightbog}

For attractive particle interaction the sign of the $ |\Psi|^4$ term in \eqref{hamilton0} changes, and a time independent solution of the corresponding GPE in a frame moving with speed $ \beta$ 
\begin{equation}
\label{gpeatt}
i \partial_t \psi(x,t) = \left( - \frac{1}{2} \partial_x^2-| \psi|^2- \mu -i \beta \partial_x \right) \psi(x,t)
\end{equation}
 is the bright soliton
\begin{equation}
\label{bright}
\psi_b(x)= e^{-i \beta (x-x_0)} \kappa \sech \kappa (x-x_0).
\end{equation}
with the chemical potential given by $ \mu=-( \kappa^2 + \beta^2)/2$. The linearization around this bright soliton background analogous to Section \ref{bogoliubov} leads to the BdGE for the mode functions
\begin{equation}
\label{bdgbright}
\begin{split}
m_ \eta \Omega_{ \eta}^2 S_ \eta &= H_b R_ \eta - \psi_b^2 R_ \eta ^ \ast \\
m_ \eta ^{-1} R_ \eta &= H_b S_ \eta +\psi_b^2 S_ \eta \\
H_b & \equiv - \frac{1}{2} \partial_x^2-2 | \psi_b|^2 -\mu - i \beta \partial_x \; .
\end{split}
\end{equation}
Notice the sign changes in \eqref{gpeatt} and \eqref{bdgbright} due to the sign change of the particle interaction. Abbreviating $ \tilde{x} =x-x_0$ the BdGE for unit mass have the non-zero frequency solutions
 \begin{equation}
\label{brightpert}
\begin{split}
R_ \eta &=( \kappa \tanh \kappa \tilde{x} \, \partial_x + \frac{k^2-\kappa^2}{2} + \kappa^2 \sech^2 \! \kappa \tilde{x})  d(\tilde{x}) \\
 S_ \eta &= (\kappa \tanh \kappa \tilde{x} \, \partial_x+ \frac{1}{2}(k^2- \kappa^2) ) \frac{d(\tilde{x})}{k^2+\kappa^2}
\\
d(x) &=e^{-i \beta x} \left\{ \begin{array}{r} \sin kx \\  \cos kx \end{array} \right. \; ,
\end{split} 
 \end{equation} 
 where both $\sin kx$ and $ \cos kx$ are possible for $d(x)$. The non-zero frequency spectrum is given by $ \Omega_{ k} =k^2/2+ | \mu|$, which reveals the macroscopic energy gap $ (\kappa^2+ \beta^2)/2$ between the lowest lying zero energy soliton translational mode (of positive mass)
 \begin{equation}
 \begin{bmatrix} R_{bz} \\ S_{bz} \end{bmatrix} \equiv \frac{1}{\sqrt{2 \kappa}} \begin{bmatrix} -(i \beta+ \partial_x) \psi_b\\ i \partial_ {\beta} \psi_b \end{bmatrix}
 \end{equation}
and the non-zero frequency modes \eqref{brightpert}.

 \end{appendix}
\bibliographystyle{apsrev}

\end{document}